\documentclass[12pt,preprint]{aastex}

\slugcomment{}

\begin{document}

\title{Testing the binary black hole paradigm through the Fe K$\alpha$
line profile: application to 3C~273}

\author{Diego F. Torres\footnote{Lawrence Livermore National
Laboratory, 7000 East Ave., L-413, Livermore, CA 94550, USA.
E-mail: dtorres@igpp.ucllnl.org}, Gustavo E.
Romero\footnote{Instituto Argentino de Radioastronom\'{\i}a (IAR),
C.C.\ 5, 1894 Villa Elisa, Argentina. E-mail:
romero@irma.iar.unlp.edu.ar}, Xavier Barcons\footnote{Instituto de
F\'{\i}sica de Cantabria (CSIC-UC), 39005 Santander, Spain.
E-mail: barcons@ifca.unican.es}, Youjun Lu\footnote{Canadian
Institute for Theoretical Astrophysics, University of Toronto,
Canada. and Center for Astrophysics, University of Sci. \&
Technology of China, Hefei, Anhui 230026, P. R. China. E-mail:
luyj@cita.utoronto.ca } }

\begin{abstract}
  We propose the study of long-term
  variations in the Fe K$\alpha$ line profile as a discriminator for
  binary black hole (BH) systems. The existence of a binary BH in the center of an active
  galaxy will produce a particular signature in the evolution of the
  line profile,  as a result of disk precession.  This signature
  is a periodic variation of the position of the blue edge of the
  profile, accompanied by periodic change of its intensity.
  We show that detection of the former is clearly within the observational
  capabilities of planned X-ray missions.  Detecting a periodic
  variation of line profiles would provide the first direct evidence
  for precessing discs in active galactic nuclei, as opposed to the
  existing evidence supporting only the precession of jets. We apply
  these ideas to 3C~273.

\end{abstract}

\section{Introduction}

The fact that many galaxies harbor supermassive BHs (SBHs) and
that galaxies often merge imply that massive binary systems might
exist at the core of some extragalactic objects (e.g. Begelman et
al.\ 1980; Valtaoja et al.\ 1989; Yu 2002). These systems are
expected to be stable for timescales of the order of the Hubble
time if they are not subject to external perturbations (Valtaoja
et al.\ 1989). SBH binaries can manifest themselves in Active
Galactic Nuclei (AGNs) through remarkable periodic signals in the
lightcurves at different wavelengths (e.g. Sillanp\"a\"a et al.\
1988; Fan et al.\ 2001), the presence of wiggling and precessing
jets (e.g. Roos 1988, Kaastra \& Roos 1992), or periodic
morphological changes at milliarcsec scales (e.g. Abraham \&
Romero 1999). Here we propose an alternative test for
binary BHs in AGNs based on high sensitivity X-ray spectroscopy of
the Fe K$\alpha$ line.

\section{Jet and disc precession in 3C~273}

3C~273 was first identified as a quasar at redshift $z=0.158$
by Schmidt (1963). Detailed observations of this object have been
performed for almost 40 years, including simultaneous measurements
of its spectrum from radio to $\gamma$-ray energies (e.g. Lichti
et al.\ 1995) and multiwavelength variability studies  (e.g. von
Montigny et al.\ 1997). At VLBI scales the source presents a
one-sided jet that extends up to $\sim 70$ milliarcsec from the
core at 5 GHz (Zensus et al.\ 1988). The absence of a counterjet
and the superluminal speeds of different components suggest a
small angle with respect to the line of sight. Zensus et al.\
(1988) estimate that this angle should be smaller than
$16^{\circ}h$, where $h$ is such that the Hubble constant is
$H_0=100h$ km s$^{-1}$ Mpc$^{-1}$.

Abraham \& Romero (1999) have used all available VLBI data of
superluminal components in 3C~273 to determine the kinematic
evolution of the inner jet. The different velocities and position
angles at the ejection time of the various well-monitored
components are consistent with the existence of a precessing inner
jet in the quasar. Fits of the VLBI data suggest a period of
precession of $\sim 16$ years in the observer's frame. For $h=0.7$
the jet would have a bulk Lorentz factor $\Gamma\sim10.8$ and
precess within a cone of half-opening angle $\sim3.9^{\circ}$,
forming a mean viewing angle of $\sim 10^{\circ}$. The Doppler
factor should vary between 2.8 and 9.4 because of the changing
viewing conditions, resulting in periodic variations in the
lightcurve, as have been found by Fan et al.\ (2001) at optical
wavelengths. Romero et al.\ (2000) have explored the dynamical
origin of the jet precession, suggesting that it might be induced
by the precession of the accretion disk, which, in turn, would be
the result of the external torque of a companion supermassive
object. The recent detection of new superluminal components in
3C~273 shows ejection angles that are in good agreement with the
predictions of the binary BH model (Krichbaum et al.\ 2001).

The precession of the accretion disk in 3C~273 will also affect
the intensity and shape of the line profiles as seen in the
observer's frame. The Fe K$\alpha$ line, for instance,
has been recently detected by Yaqoob \& Serlemitsos
(2000) using {\it ASCA} and {\it RXTE} observations. The line is
as broad as those observed in Seyfert 1 galaxies, with a mean
Gaussian width of $0.8\pm 0.3$ keV, corresponding to a FWHM of
$\sim0.3\pm0.1c$. The intrinsic 2--10 keV luminosity is in the
range $\sim (0.8$--$2.0)\times 10^{46}$ erg s$^{-1}$.

\section{A line profile diagnosis}

\begin{figure*}[t]
\centering
\includegraphics[width=14cm,height=6cm]{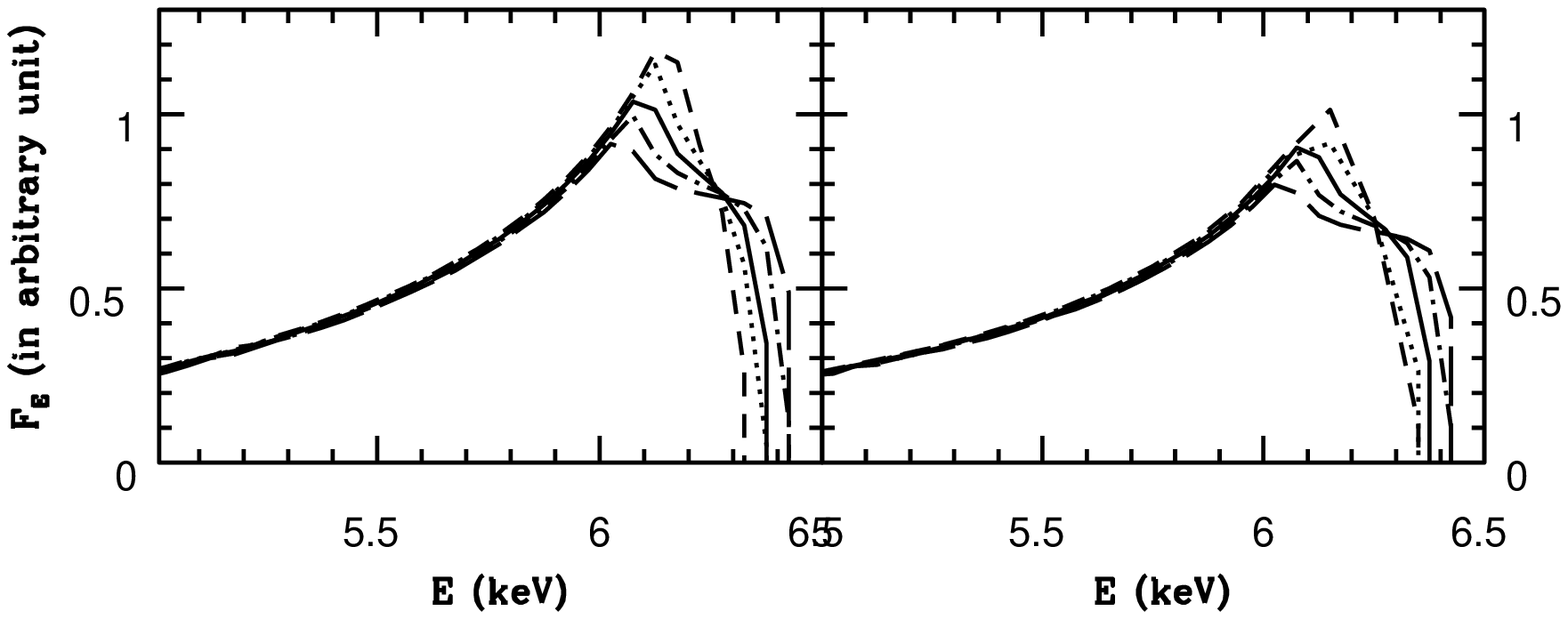}
\includegraphics[width=14cm,height=6cm]{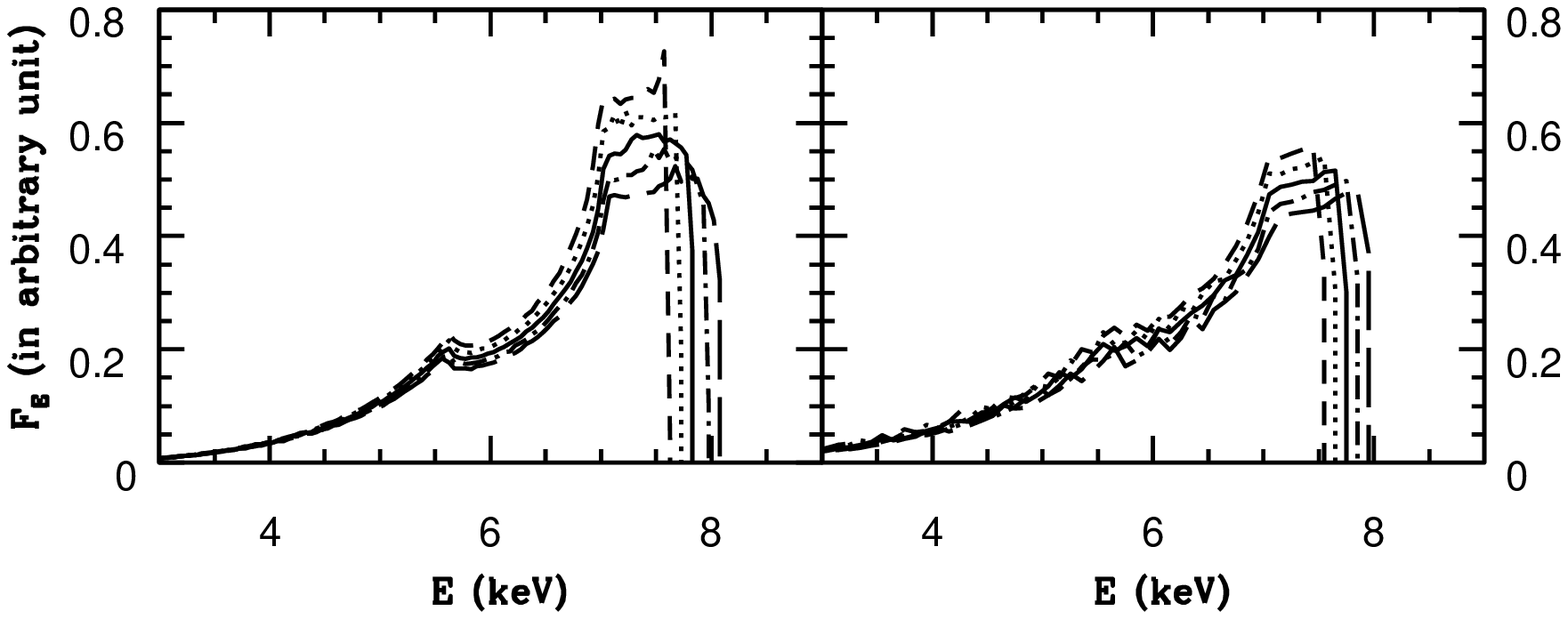}
\caption{Top: Evolution of the Fe K$\alpha$ profile for 3C~273
(right panel: Kerr BH with $a/M=0.998$; left panel: Schwarzschild
BH). In each panel, the lines from right to left --at the blue
edge-- present the prediction for different inclination angles
along the 16-yr precession period: $\theta_{\rm i}=14^\circ$,
12$^\circ$, 10$^\circ$, 8$^\circ$, and 6$^\circ$. The blue edge of
the Fe K$\alpha$ line changes periodically. Bottom: Similar to top
panel, but now for an average inclination angle of
$\langle\theta_i\rangle =60 ^\circ$. Again, from right to left at
the blue edge: $\theta_{\rm i}=64^\circ$, 62$^\circ$, 60$^\circ$,
58$^\circ$, and 56$^\circ$. }
\end{figure*}

We use the ray-tracing technique to calculate the profile of the
Fe K$\alpha$ line (Fabian et al.\ 1989; Yu \& Lu 2000, Lu \& Yu 2001 and
references therein). The parameters of the simulation are the
inclination angle of the normal relative to the line of sight of
the observer ($\theta_{\rm obs}$), the inner disk radius ($r_{\rm
in}$, assumed constant during the period in which the line
variation is studied), the outer disk radius ($r_{\rm out}$), and
the power-law index characterizing the surface emissivity of line
photons, which we assume is proportional to $r^{-q}$. We also
assume that the disk extends down to the innermost marginal stable
orbit ($r_{\rm in}=r_{\rm ms}$; $r_{\rm in}=6r_{\rm g}$ for a
Schwarzschild BH and $r_{\rm in} =1.23r_{\rm g}$ for a Kerr BH
with spin $a/M=0.998$), and that the outer part of the precessing
disk is at $r_{\rm out}=160 r_{\rm g}$, where $r_{\rm g}=GM/c^2$
is the gravitational radius. We adopt $q=2.5$ for our
calculations, which is the average emissivity exponent index
obtained from the fitting of the Fe K$\alpha$ line for a sample of
AGNs observed by {\it ASCA} (Nandra et al.\ 1997). Since $q$ is a
positive number, the line emission is dominated by the innermost
region with sizes of a few to tens of gravitational radii. The
line profile is not sensitive to the chosen outer radius. Note
that
if the line profile is different, for example due to disk
ionization, the overall arguments will continue to apply under the
only assumption that the intrinsic line emission mechanism does
not change along the timescale relevant for disk precession.

The line profile depends on the mass through the combination
$M/r$; it is determined by the Doppler shift due to Keplerian
motion of the disk material and the general relativistic
effects.\footnote{If we choose the same inner and outer radius of
the emission region in units of the Schwarzschild radius, the same
inclination angle and the same emissivity law, we would predict
the same line profile whatever the BH mass is.} Romero et al.'s
(2000) dynamical modelling of 3C~273 explains the jet precession
by a tidal perturbation on the accretion disk due to a secondary
BH. Different BH masses are obtained with somewhat different
models. However, for the observational test we are proposing, it
does not matter what the BH mass is, since we only need to know
the changes of the disk inclination angle in order to compute the
profile evolution.

Abraham \& Romero (1999) have found that the half opening-angle of
the precessing jet (assumed to be normal to the disk) in 3C~273 is
about 3.9$^\circ$, thus the disk inclination angle should slowly
change periodically from $\theta_{i0}$ to $\theta_{i0}+8^\circ$.
The angle $\theta_{i}$ is not well-known. We adopt here an average
value $\langle\theta_i\rangle =10 ^\circ$ in accordance with the
best fit for $h=0.7$ by Abraham \& Romero (1999). This value is in
good agreement with the estimates by Zensus et al.\ (1988), which
are derived from jet observations. However, both the model of the
UV band spectrum in 3C~273 by Kriss et al.\ (1999) and the
spectral fits of the Fe K$\alpha$ in this object by Yaqoob \&
Serlemitsos (2000) suggest a far larger value of $\sim 60^\circ$.
This value seems to contradict the fact that 3C~273 is a
blazar-like object and a superluminal source (a behavior that
requires a close alignment between the jet and the line of sight,
i.e. a nearly face-on disk). Notwithstanding, we have also
calculated the Fe K$\alpha$ profile evolution for a disk with
$\langle\theta_i\rangle =60 ^\circ$.
We are assuming that only one of the BHs has a significant
accretion disc. This may well not be the case, and a larger
inclination angle
could result in the --probably unlikely-- case of the jetless BH
producing the profile.

The results of our calculations are shown in Fig.~1:
the blue edge of the Fe K$\alpha$ line moves down to a lower
energy as the inclination angle decreases from 10$^\circ$ to
6$^\circ$, then moves back as the inclination angle increases, and
finally extends to a higher energy when the inclination angle goes
to 14$^\circ$. The total shift in energy of the peak of the line,
along a span of 8 years, is 0.1 keV (at $\sim 6$ keV). On the same
timescale, the variation of the intensity is $\Delta I/I\sim
33$\%. The same kind of behavior can be seen in the bottom panel
of Fig~1: a shift in energy of 0.4 keV at
$\sim7.5$ keV and an intensity change of $\Delta I/I\sim 26$\%.
However, note that if the relation between continuum and line
emission is not exactly known, we can not put much emphasis in
observing line intensity variations as a signature of disc
precession.

\section{Feasibility}

Although the Fe K$\alpha$ emission line in 3C~273 is relatively
weak, and the effect produced by the disc precession is small,
this would be detectable by a high-throughput high-spectral
resolution mission such as {\it XEUS\/} or {\it
Constellation-X\/}. {\it XEUS} is a mission currently under study
by the European Space Agency (ESA) and Japan, with an expected
launch around 2015. See the {\it XEUS\/} baseline
mission profile in Bleeker et al. (2000) and the {\it
XEUS\/} science case in Arnaud et al. (2000).
For the purposes of observing the Fe line at high resolution, the
most appropriate of the foreseen instruments is the Narrow-Field 2
instrument, which is based on Transition Edge Sensors (see Hoevers
\& Verhoeve 2003). We have used the response matrices given in the
{\it XEUS\/} web page at ESA.

Simulations have been performed using the parameters found by
Yaqoob \& Serlemitsos (2000).  The underlying continuum is
described by a single  power law with $\Gamma=1.6$, to which an Fe
line, according to the above profiles, has been added.  The
equivalent width of the emission line (for which there are
reported values from 100 to 300 eV) is assumed to be 200 eV.
Galactic photoelectric absorption with column density $N_H=1.79
\times 10^{20}$ cm$^{-2}$  has been applied. The total unabsorbed
2--10 keV flux is assumed to be $8.4\times 10^{-11}\, {\rm erg}\,
{\rm cm}^{-2}\, {\rm s}^{-1}$, which corresponds to a luminosity
at $z=0.158$ of $6.25 \times 10^{45}\, {\rm erg}\, {\rm s}^{-1}$
for a cosmology ($H_0=65\, {\rm km}\, {\rm
s}^{-1}\, {\rm Mpc}^{-1}$, $\Omega_m=0.3$ and
$\Omega_{\Lambda}=0.7$).

Simulations have been conducted with {\tt xspec} (Arnaud 1996),
version 11.2.0.  3C~273 would have a count rate of  $\sim 10^4$\
cts/s in the {\it XEUS\/} Narrow-field  2 detector, but since the
specification is that it should be able to read up to 30 kHz,
pile-up should not pose an unsolvable  problem for this
observation. If necessary, a thick filter suppressing soft X-ray
photons (which are not interesting for this project) will solve
this problem. All simulations were done for 100 ks and without any
background. Counts were grouped on bins containing at least one
million counts. We simulated one spectrum for each of the curves
in Fig. 1  and fit a power-law only to the range 2--4 keV and
8--12 keV. Fig. 2 --top panels-- (for a Schwarzschild and a Kerr
case) display the ratio of the data to the absorbed power law
model. The differences in the blue edge of the emission lines are
readily apparent.

\begin{figure*}[t]
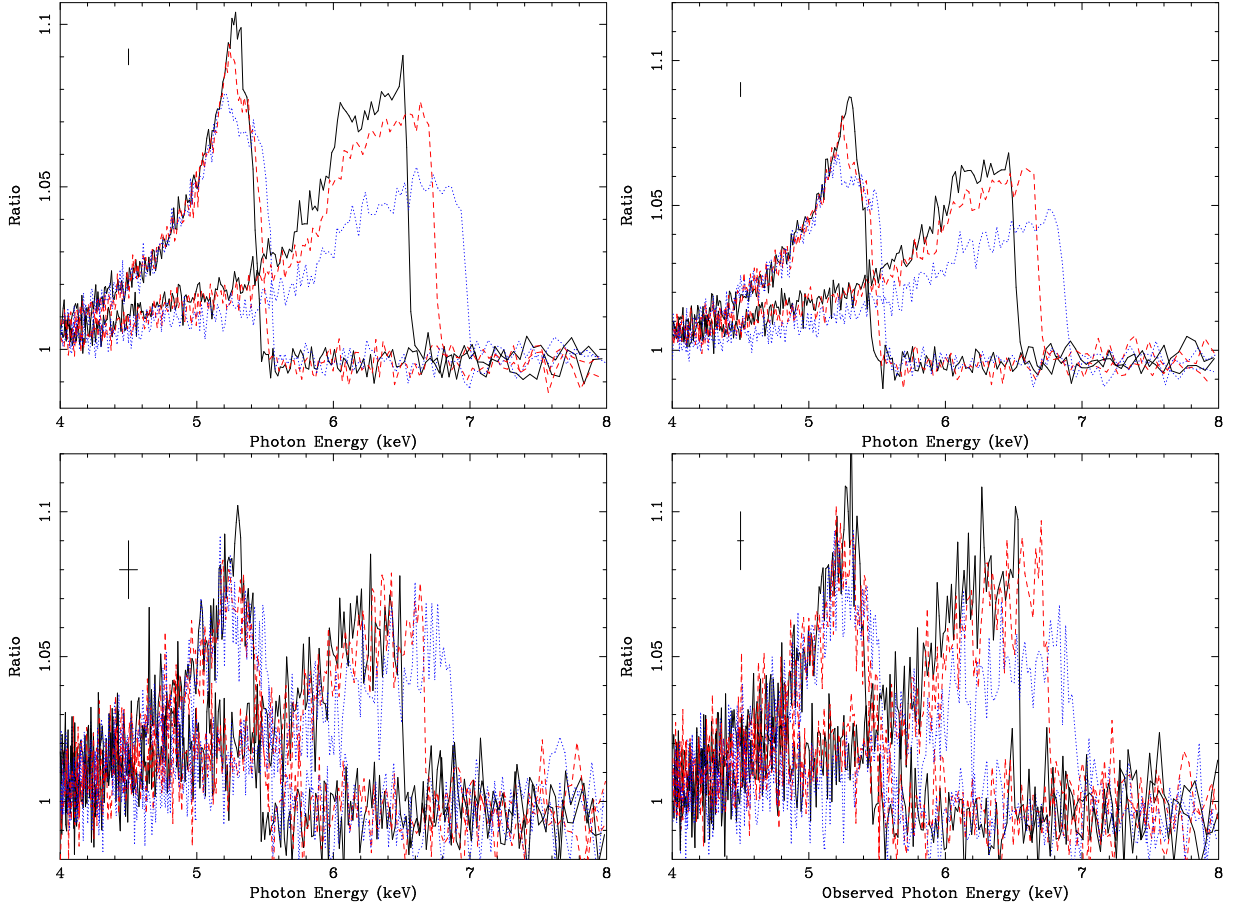

\centering
\includegraphics[angle=-90,width=8cm]{f2a.ps}
\includegraphics[angle=-90,width=8cm]{f2b.ps}
\includegraphics[angle=-90,width=8cm]{f2c.ps}
\includegraphics[angle=-90,width=8cm]{f2d.ps}
\caption{Top Left: The simulated Fe K$\alpha$ line profile of
3C~273 for a Schwarzschild BH as seen through the NFI2 detector of
{\it XEUS}. Disk inclination angles are 6$^{\circ}$ (continuous),
10$^{\circ}$ (dashed) and 14$^{\circ}$ (dotted) for the leftmost
profiles and 56$^{\circ}$ (continuous), 60$^{\circ}$ (dashed) and
64$^{\circ}$ (dotted) for the rightmost profiles. The vertical
mark on the top left is the size of the typical statistical error
bar in each bin. Note that the lines are redshifted with respect
to the emission profiles of Fig. 1. Top Right: idem for a Kerr BH.
Bottom Left and Right: Idem above but considering the calorimeter
with 2 eV resolution of {\it Constellation-X}.}
\end{figure*}

Similarly, we have also checked whether the {\it Constellation-X}
mission would also be able to detect such differences with the ``2
eV'' calorimeter (i.e.  a calorimeter with 2 eV spectral
resolution) and with a similar integration time. For review
articles on the science and instruments of {\it Constellation-X}
see Tr\"umper and Tananbaum (2003). Although its effective area
would be a factor of several smaller than that of {\it XEUS\/} at
these energies, the effect is still detectable (although not so
visually obvious) with a similar exposure time, see Fig. 2
--bottom panels--. The only difference with respect to the {\it
XEUS} simulations presented in the top panels of the same figure
is that counts are grouped in bins containing $1\times 10^5$
counts as opposed to $1\times 10^6$ count bins for the former
case.

To further illustrate the capabilities of both missions to detect
the drift in the blue edge of the lines, we have performed a
traditional fitting exercise to the simulated X-ray spectra. We
have used model lines with Kerr profiles and various inclination
angles, and then fitted the grouped spectra in the 2--12 keV band
according to the above-explained grouping criteria.  The fitting
model is chosen as the sum of a power law, a maximally rotating
Kerr, and a Schwarzschild BH line profiles, in order to partially
mimic our likely limited knowledge on the BH spin.  The
inclination of the disk is left free but the same for both line
components, whereas the intensities of both line components are
independently left as free parameters too. Models are generated by
{\tt xspec} by linearly interpolating inclination angles over our
$2^{\circ}$ step models. Since we are interested just in showing
further evidence for the detectability of the periodic variation
of the blue edge of the line,
this approach is enough for illustrative purposes. Results of the
fitting exercise are presented in Table~1. For small inclination
angles, the proposed {\it Constellation-X} observations would
deliver 90\% confidence measurements of the angle within $\sim \pm
0.7^{\circ}$, while variations of $\sim 1^{\circ}$ per year are
expected. The situation is much more favorable with {\it XEUS},
where the inclination angle can be measured with a 90\% confidence
within $\pm 0.3^{\circ}$.  In the high inclination situation, the
errors are much smaller and the blue edge variations more easily
detectable for both telescopes.

\begin{table}
\caption{Fitting exercise to simulated spectra with
an   Fe Kerr line profile with inclination angle $\theta_{in}$
(all angles in degrees).  Errors are 90\% confidence level for a
single interesting parameter.}
\begin{center} \begin{tabular}{r r r} \hline
$\theta_{in}$ & ${\theta_{fit}}$ {\it [CON-X]} &  $\theta_{fit}$ {\it [XEUS]}\\
\hline
8.0 & $7.4^{+0.7}_{-0.6}$ & $7.9^{+0.2}_{-0.2}$\\
10.0 & $9.9^{+0.6}_{-0.7}$ & $10.0^{+0.4}_{-0.3}$\\
12.0 & $11.8^{+0.7}_{-0.8}$ & $11.8^{+0.3}_{-0.2}$\\ \hline
58.0 & $57.6^{+0.4}_{-0.2}$ & $57.9^{+0.2}_{-0.1}$\\
60.0 & $59.9^{+0.3}_{-0.2}$ & $60.0^{+0.1}_{-0.1}$\\
62.0 & $62.1^{+0.4}_{-0.4}$ & $61.9^{+0.2}_{-0.1}$\\
\hline
\end{tabular}
\end{center}
\end{table}

Calibration, as in any other X-ray
observation, needs to be known to high precision.  In order to
detect the Fe line itself the effective area needs to be known to
better than 10\%, whereas to detect differences between the blue
edge of the various profiles shown in Fig. 2, it needs to be
better than a few per cent. Furthermore, the energy scale needs to
be known to within less than 10 eV (the drift corresponds to $\sim
20\, {\rm eV}$ per degree). It is expected that both the {\it
Constellation-X} and {\it XEUS} high spectral resolution
instruments will have a well calibrated and monitored energy scale
to a level at least commensurate with the spectral resolution of a
few eV. Under these assumptions, the variation of the blue edge of
the line can be precisely tracked as it changes in time because of
the disc precession, ultimately probing the innermost regions of
the galaxy and supporting or disproving a binary BH interpretation
for 3C~273.

\section{Discussion}

Short term variability (on time scales of about $10^4$s) of both
the intensity and shape profile of the Fe K$\alpha$ line are an
important feature revealed by observations of some Seyfert
galaxies (Iwasawa et al.\ 1996; Nandra et al.\ 1999). This
variability may be produced by X-ray flares above the accretion
disk. There could also exist similar variability in the profile of
the Fe K$\alpha$ line in 3C~273. However, this component, if
present, will be chaotic, not producing a periodic long-term
variation as predicted here for the precession.
It is also true that the emissivity law could be a function of
azimuthal angle. However, there is no reliable model on
the production mechanism for the assumed X-ray flares that gives
their possible distribution or the dependence of line emissivity
on the azimuthal angle that we are aware of. One could {\it
arbitrarily} assume a line emissivity dependence on azimuthal
angle, but with current information this would be just as good as
assuming a uniform distribution.

The essential point of this research is not to
produce a fully testable model for the Fe K$\alpha$ line, but
rather to look for an unambiguous signature of disk precession in
the line profile evolution. The drift of the blue edge of the line
(averaged or properly sampled over, at least, an orbital period)
can only be due to disk precession if, whatever the emission
mechanism is (including its possible azimuthal dependence), it
does not change in time. The relevant point here is whether or not
we sample an azimuth-averaged situation, since sampling different
phases could lead to distinct line profiles.  But this can be
reliably done, whatever the emission properties, with a properly
tailored observing strategy.
Each visit to the target should represent an azimuth-averaged
observation. 
For 3C273, with mass
$\sim 10^9 M_\odot$, the orbital timescale at 10 $r_g$ is a few
times $10^5$ s, with the observed line photons dominated by those
coming from the inner region (with $R \sim 10 \, r_g$ or less). An
exposure of several days on 3C273 would then average any putative
dependence of the line emissivity law on azimuthal angle. Also, a
few 10--50 ks snapshots spread over a few orbital periods can be
equally used for averaging any inhomogeneity over. Note that 3C273
is a very likely calibration target for any X-ray mission. This
means that these rather demanding observations could be conducted
anyway.
Therefore, several observations over a total span of the order of
the precession period ($\sim 16$ yr) might reveal the systematic
change of the line profile associated with the presence of a SBH
companion. The detection of the predicted changes in the line
profile would provide direct evidence of disk precession, which
can be considered as even stronger evidence for a massive
companion than the evidence from the jet precession (since
instabilities in the jet can in some cases be
responsible for twisting, e.g. Romero 1995).

\section*{Acknowledgments}

The  work of DFT was performed under the auspices of the U.S.
DOE--NNSA by U. of California LLNL under contract No.
W-7405-Eng-48. GER is supported by PICT 03-04881, PIP 0438/98, and
F. Antorchas. XB acknowledges partial financial support
by the Spanish Ministerio de Ciencia y Tecnolog\'\i a, through
project AYA2000-1690. DFT acknowledges C. Mauche for insightful
comments. Two anonymous Referees are also acknowledged.

\end{document}